\documentclass[useAMS,usenatbib,referee]{mn2e}

\title[Studies of dissipative standing shock waves around black holes]
  {Studies of dissipative standing shock waves around black holes}
\author[Santabrata Das, Sandip K. Chakrabarti and Soumen Mondal]
  {Santabrata Das\thanks{sbdas@canopus.cnu.ac.kr,sbdas@iitg.ernet.in}$^{1, 2}$,
  Sandip K. Chakrabarti\thanks{chakraba@bose.res.in} $^{{3, 4}}$
  and Soumen Mondal\thanks{soumen@bose.res.in}$^{5, 1}$
  \newauthor % starts a new line in the
  \\
  $^1$Korea Astronomy and Space Science Institute,
      61-1, Hwaam Dong, Yuseong-Gu,
      Daejeon 305 348, South Korea.\\
  {$^2$Indian Institute of Technology Guwahati,
      Guwahati, 781039, Assam, India.}\\
  ${^3}$ S. N. Bose National centre for Basic Sciences,
  JD-Block, Sector III, Salt Lake, Kolkata, 700 098, India.\\
  ${^4}$ Centre for Space Physics, Chalantika 43, Garia Station Rd.,
  Kolkata 700084, India\\
  ${^5}$ Ramakrishana Mission Residential College,
   Narendrapur, Kolkata 103, India.}

\date{\today}
\pagerange{\pageref{firstpage}--\pageref{lastpage}} \pubyear{2002}

\def\LaTeX{L\kern-.36em\raise.3ex\hbox{a}\kern-.15em
    T\kern-.1667em\lower.7ex\hbox{E}\kern-.125emX}

\usepackage{graphicx}
\input psfig.tex

\def\lsim{\mathrel{\hbox{\rlap{\hbox{\lower4pt\hbox{$\sim$}}}\hbox{$<$}}}}
\def\gsim{\mathrel{\hbox{\rlap{\hbox{\lower4pt\hbox{$\sim$}}}\hbox{$>$}}}}
\def \simeq{\lower.3ex\hbox{$\; \buildrel \sim \over - \;$}}

\begin{document}

\label{firstpage}

\maketitle

\begin{abstract}
We investigate the dynamical structure of advective accretion flow
around stationary as well as rotating black holes. For a suitable
choice of input parameters, such as, accretion rate ($\dot {\cal M}$) 
and angular momentum ($\lambda$), global accretion solution may 
include a shock wave. 
The post shock flow is located at few tens of Schwarzchild radius
and it is generally very hot and dense. This successfully mimics the
so called Compton cloud which is believed to be responsible for emitting hard
radiations. Due to the radiative loss,
a significant energy from the accreting matter is removed and
the shock moves forward towards the black hole in order to
maintain the pressure balance across it. 
We identify the effective area of the parameter space 
($\dot {\cal M} - \lambda$) which allows accretion flows to
have some energy dissipation at the shock $(\Delta {\cal E})$.
As the dissipation is increased, the parameter space is reduced and finally
disappears when the dissipation is reached its critical value. 
The dissipation
has a profound effect on the dynamics of post-shock flow. By moving forward, 
an unstable shock whose oscillation causes Quasi-Periodic Oscillations (QPOs) 
in the emitted radiation, will produce oscillations of high frequency. 
Such an evolution of  QPOs has been observed in several black hole 
candidates during their outbursts.
 
\end{abstract}


\section{Introduction}	

In a significant work on the prospect of shock formation in an accretion disk 
around a black hole, \citet{cd04} 
showed that in order to have a stable shock, the viscous dissipation 
inside a flow must have an upper limit, beyond which the Rankine-Hugoniot 
conditions cannot be satisfied. In \citet{dc04} and \citet{d07}, the
bremsstrahlung and synchrotron cooling
were also added to dissipate away the heat generated from viscous dissipation.
However, it is well known that the post-shock region emits the hard X-rays
in a black hole candidate \citep{ct95}
and some amount of energy is lost through radiation from the post-shock region.
This radiative loss primarily comes from the thermal energy of the 
flow and takes place via thermal Comptonization. In a self-consistent
shock condition, this radiative loss must also be incorporated.
In the present paper, we quantitatively show how the energy loss 
at the shock
affects the location of the shock itself around stationary as well as
rotating black holes. As energy dissipation is increased, the post-shock
flow pressure gets reduced causing the shock front to come closer to the
black hole in order to maintain the pressure balance across it. Accordingly,
the dynamical properties of standing shock waves
would directly be related to the amount of energy discharge from the
post-shock flow. In addition, the mass outflow rate which is believed to be 
generated from the post-shock region \citep{c99,dcnc01,dc08}, would also be
affected by the energy discharge at the shock location. Therefore, it is
pertinent to understand the response of energy dissipation on the formation
of standing shock wave. In this paper, we precisely do this.

The plan of our paper is the following: in the next Section, 
we present the equations governing the flow and the procedure
we adopted to solve these equations. In \S 3, we show how the 
Rankine-Hugoniot conditions at the shocks must be modified when energy
dissipation is present. In \S 4, we show the results of our computations.
Finally in \S 5, we present concluding remarks.

\section{Governing equations and sonic point analysis}
We start with a steady, thin, rotating, axisymmetric,
accretion flow around black holes. We assume smaller accretion rates, so that 
the flow radiatively inefficient and behaves essentially as a 
constant energy flow as in Chakrabarti (1989). We assume a polytropic
equation of 
state for the accreting matter, $ P = K \rho^\gamma$, where, $P$ and 
$\rho$ are the isotropic pressure and the matter density, respectively, 
$\gamma$ is the adiabatic exponent considered to be constant throughout 
the flow and $K$ is a constant which measures the entropy of the flow
and can change only at the shock. Since we ignore viscous dissipation
the angular momentum of the flow $\lambda \equiv x \vartheta_{\theta}$ 
is also constant everywhere. However, we assume that the main dissipation is
concentrated in the immediate vicinity of the post-shock flow, which would 
be the case if the thermal Comptonization is the dominant process.
The flow height is determined 
from the condition of being in equilibrium in a direction perpendicular
to the equatorial plane. Flow equations are made dimensionless considering 
unit of length, time and the mass as $GM_{BH}/c^2$, $GM_{BH}/c^3$ 
and $M_{BH}$ respectively, where $G$ is the gravitational constant, 
$M_{BH}$ is the mass of the black hole and $c$ is the velocity of light. 

In the steady state, the dimensionless energy equation at the disk 
equatorial plane is given by \citep{c89},
$$
{\cal E}=\frac{1}{2}\vartheta^2+\frac {a^2}{\gamma-1}
+\Phi,          
\eqno(1)
$$
where, ${\cal E}$ is the specific energy, $\vartheta$ is the radial 
velocity and $a$ is the adiabatic sound speed defined as
$a=\sqrt {\gamma P/\rho}$. Here, effective potential due to black
hole is denoted by $\Phi$. In the present study, the
effect of gravity is taken care of by two different potentials. To represent
Schwarzschild black hole, we use Paczy{\'n}ski-Wiita \citep{pw80} potential
$(\Phi_{PW})$ and for the Kerr black hole, we consider pseudo-Kerr 
potential $(\Phi_{PK})$ introduced by \citet{skcsm06}. It has been adequately 
shown that these potentials accurately mimic not only the geometry of the 
space-time, but also the dynamics of the flow. In fact the error for not using 
full general relativistic treatment has been shown to be at the most a few 
percent \citep{skcsm06}. The expressions for Paczy{\'n}ski-Wiita and
pseudo-Kerr effective potential are respectively given by,

$$
\Phi_{PW}=\frac{\lambda^2}{2x^2}-\frac{1}{2(R-1)}
$$
and

$$
\Phi_{PK}=-\frac{B+\sqrt{B^2-4AC}}{2A}
$$
where, 

$$
A=\frac{\alpha^2 \lambda^2}{2x^2},
$$

$$
B=-1 + \frac{\alpha^2 \omega \lambda R^2}{x^2} 
+\frac{2a_k\lambda}{R^2 x}
$$

$$
C=1-\frac{1}{R-x_0}+\frac{2a_k\omega}{x}
+\frac{\alpha^2 \omega^2 R^4}{2x^2}.
$$
Here, $x$ and $R$ represent the cylindrical and spherical radial distance 
from the black hole when the black hole itself is considered to be located 
at the origin of the coordinate system. Here, 
$x_0=(0.04+0.97a_k+0.085a_k^2)/2$, $\omega=2a_k/(x^3+a^2_k x+2a^2_k)$ 
and $\alpha^2=(x^2-2x+a^2_k)/(x^2+a_k^2+2a^2_k/x)$, $\alpha$ is the 
red shift factor. $a_k$ represents the black hole rotation parameter
defined as the specific spin angular momentum of the black hole. 

The mass flux conservation equation in the steady state apart from 
the geometric factor is given by,
$$
{\dot M}=\vartheta \rho x h(x),
\eqno(2)
$$
where, ${\dot M}$ is the mass accretion rate considered to be constant,  
and $h(x)$ represents the half-thickness of the flow \citep{c89} 
which is expressed as,
$$
h(x)=a \sqrt{\frac{x}{\gamma \Phi^{'}_R}}.
\eqno(3)
$$
Here, $\Phi^{'}_R=\left(\frac{\partial \Phi}{\partial R}\right)_{z<<x}$ 
and $z$ is the vertical height in the cylindrical 
co-ordinate system where $R=\sqrt{x^2+z^2}$. By using the polytropic
equation of state and the definition of the adiabatic sound speed, we get the
equation for the so-called `entropy accretion rate' \citep{c89} as:  
$$
{\dot {\cal M}}=\vartheta a^\nu \sqrt{\frac{x^3}{\gamma \Phi^{'}_R}} 
\eqno(4)
$$
where, $\nu=(\gamma+1)/(\gamma-1)$. 

In order to form shocks, accretion flow must be supersonic at 
some point, $i.e.$, stationary flow must pass through a 
sonic point. We derive the sonic point conditions following the 
standard sonic point analysis \citep{c89}. We, therefore, differentiate 
Eqs. (1) and (4) with respect to $x$ and eliminate terms that involve
derivatives of $a$ to obtain radial velocity gradient which is expressed as,
$$
\frac{d\vartheta}{dx}=\frac{N}{D},
\eqno(5)
$$
where, the numerator $N$ and the denominator $D$ are respectively given by, 
$$
N= \frac{3a^2}{x(\gamma+1)}
-\frac{d \Phi_{e}}{d x}
-\frac{a^2}{(\gamma+1)\Phi^{'}_{R}}
\frac{d\Phi^{'}_{R}}{dx}
\eqno(5a)
$$
and 
$$
D=\vartheta-\frac{2a^2}{(\gamma+1)\vartheta}.
\eqno(5b)
$$
Here, a subscript `e'  signifies that the quantity was calculated
on the equatorial plane.

In order to have the flow smooth everywhere, the numerator and the denominator 
in Eq. (5) must vanish simultaneously. These considerations allow us to obtain 
the sonic point conditions. Therefore, setting $D=0$, we have,
$$
\vartheta_c^2(x_c)=\frac{2}{(\gamma+1)}a_c^2(x_c)~~\Rightarrow~~
M_c=\sqrt{\frac{2}{\gamma+1}}.
\eqno(6a)
$$
The subscript `c'  signifies that the quantity was computed at the sonic
points. Note that the Mach number $[M(x_c)]$ at the sonic point is not
unity as in the other models since the vertical equilibrium model is
considered. The other condition comes while setting $N=0$ and is given by,
$$
a_c^2(x_c)=(\gamma+1)\left(\frac{d\Phi_e}{dx}\right)_{c}
\left[ \frac{3}{x}- \frac{1}{\Phi^{'}_{R}}\frac{d\Phi^{'}_{R}}{dx}
\right]^{-1}_{c}.
\eqno(6b)
$$
Sound speed at the sonic point can now easily be calculated from 
Eq. (6b) and it must be always positive. Flow may possess at most
three sonic points outside the black hole horizon. Among them, the closest one
from the black hole is called inner sonic point whereas the furthest
one is known as outer sonic point.  The nature of
the sonic point solely depends on the value of velocity gradient
(Eq. 5) at the sonic point. Usually, $d\vartheta/dx$ has two values
at the sonic point: one is for accretion flow and the other is for
winds. A sonic point is referred as saddle-type when both the derivatives
are real and of opposite sign. When derivatives are real and of same
sign, the sonic point is called as nodal type. If the derivatives are
complex, the sonic point is center-type. The saddle type sonic point
receives special importance as transonic flow can only pass through
it. Moreover, for standing shock transition, flow must have multiple
saddle type sonic points. We solve Eqs.(1-6) to obtain a 
full set of global flow solutions where standing shock may be present.

\section{Shock conditions and shock invariant}

In order to have a shock, the flow velocity must jump discontinuously from
super-sonic to sub-sonic branch. This discontinuous shock jump is
characterized by four flow variables, namely, shock location $(x_s)$,
radial velocity $(\vartheta)$, sound speed $(a)$ and entropy constant $(K)$
respectively and are given by,

$$
x=x_s,
\eqno(7a)
$$
$$
\Delta \vartheta=\vartheta_+(x_s)-\vartheta_-(x_s),
\eqno(7b)
$$
$$
\Delta a=a_+(x_s)-a_-(x_s),
\eqno(7c)
$$
and
$$
\Delta K=K_+-K_-,
\eqno(7d)
$$
where, the subscripts `-' and `+' denote the quantities before and
after the shock. Here, we are dealing with a quasi two dimensional
flow where pressure and density are averaged over in the vertical
direction. In this case, the Rankine-Hugoniot shock conditions
\citep{ll59} are given by,

$$
{\cal E}_+={\cal E}_- - \Delta {\cal E},
\eqno(8a)
$$
(II) the self-consistent pressure condition, 
$$
W_+ + \Sigma_+\vartheta^2_+=W_- + \Sigma_-\vartheta^2_-,
\eqno(8b)
$$
and (III) the baryon number conservation equation,
$$
{\dot M}_+={\dot M}_-.
\eqno(8c)
$$
Here, $W$ and $\Sigma$ denote the vertically averaged pressure and density
of the flow \citep{mk84}.

Because of our choice of radiatively inefficient flow, i.e., flow with
low accretion rates, the energies ${\cal E}_-$ and ${\cal E}_+$ are
nearly constant. However, in presence of dissipation at the shock,
$\Delta {\cal E}$ will be  non-zero. Since the dissipation is expected
to be mostly through thermal Comptonization, which may cool the flow
from $\sim 100$keV to $\sim 3$keV \citep {ct95}, we assume that the energy 
dissipation in the post-shock region reduces the temperature of the flow
and the loss of energy is proportional to the temperature difference between
the post-shock and the pre-shock flows similar to the well known Kirchoff's
law, i.e.,
$$
\Delta {\cal E}=\Delta {\cal E}^{'}(a_{+}^2-a_{-}^2),
$$
where, $\Delta {\cal E}^{'}$ is the proportionality constant which will
be used as a parameter. In reality, this is a function of the electron and
low energy photon contents of the accretion flow and can be calculated in
any situation \citep{ct95}. Since the thermal energy content (the final
term in Eq. 1) can be anywhere between ten to thirty percent of the rest
mass energy, a few percent of thermal energy can be easily extracted.
Below, we show that maximum dissipation for a black hole of $a \sim 0.8$
is about $6.5$\%.

\noindent(I) the energy conservation equation,

Using the governing equations and shock conditions, we compute an invariant
relation at the shock \citep{c90} which is given by,

$$
C=\frac{[M_+(3\gamma-1)+\frac{2}{M_+}]^2}
{2+(\gamma-1)M^2_+ +2(\gamma-1)\Delta {\cal E}^{'}}=
\frac{[M_-(3\gamma-1)+\frac{2}{M_-}]^2}
{2+(\gamma-1)M^2_- +2(\gamma-1)\Delta {\cal E}^{'}},
\eqno(9)
$$
where, $M$'s are the Mach numbers of the flow.

Simplifying Eq. (9) we have the following relations,
$$
M_{\pm}=\frac{-{\cal Y}\pm\sqrt{{\cal Y}^2-4{\cal X}{\cal Z}}}{2{\cal X}} ,
\eqno(10)
$$
and 
$$
M_{+}M_{-}=\frac{2}
{\sqrt{(3\gamma-1)^2-C(\gamma-1)}},
\eqno(11)
$$
where,
$$
{\cal X}=(3\gamma-1)^2-C(\gamma-1),~~
{\cal Y}=2\left[2(3\gamma-1)-C-C(\gamma-1)\Delta{\cal E^{'}}\right] ,
\ \ {\rm and} \ \
{\cal Z}=4.
$$

The flow will pass through the shock if the shock invariant condition
is satisfied (Eq. 9) at some point between the two saddle type sonic points.  

\section{Results}

\begin{figure}
\vbox{
\vskip -0.3cm
\centerline{
\psfig{figure=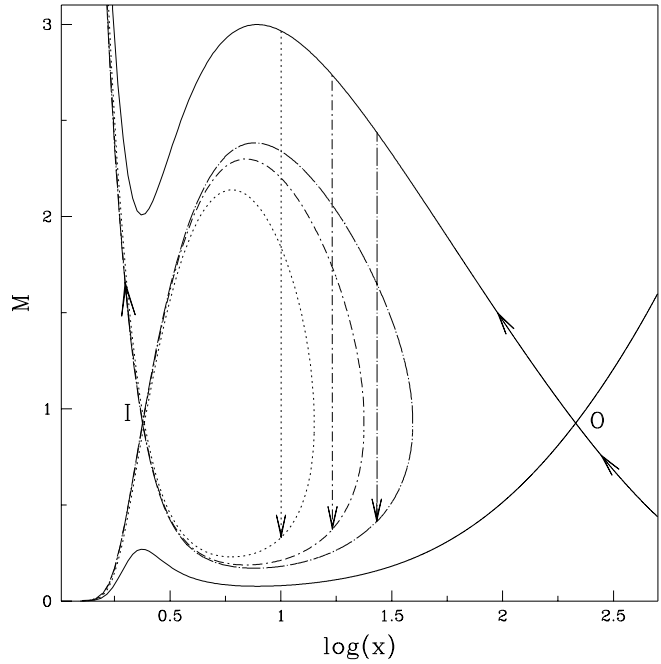,height=10truecm,width=10truecm}}
\vspace{-0.0cm}
\caption[] {
Phase space diagram of accretion flow in the M-log(x) plane. Flow 
parameters are $\lambda=3.56$ and ${\dot {\cal M}=0.08 \times 10^{-5}}$. 
The energy at outer sonic point is obtained as
${\cal E}=0.13452 \times 10^{-2}$. The vertical lines indicate the
possible shock transitions $x_s = 27.75$ (big-dashed-dot),
$x_s = 17.02$ (small-dashed-dot) and $x_s = 10.001$ (dotted)
corresponding to the energy dissipation at the shock
$\Delta {\cal E}=0.699 \times 10^{-3}$, $\Delta {\cal E}=0.37
\times 10^{-2}$ and $\Delta {\cal E} = 0.985 \times 10^{-2}$ respectively.
Shocks form closer to the black hole horizon as the energy dissipation
at the shock is enhanced.
}}
\end{figure}

\begin{figure}
\vbox{
\vskip -0.3cm
\centerline{
\psfig{figure=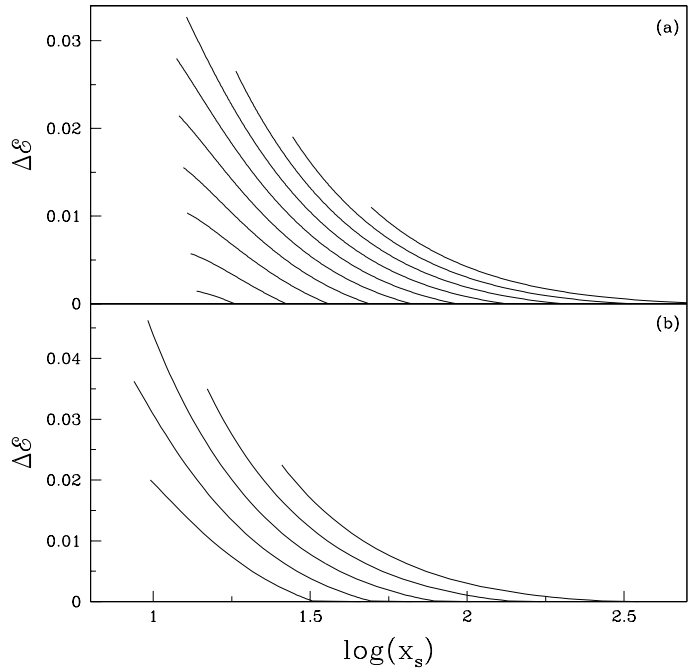,height=10truecm,width=10truecm}}
\vspace{-0.0cm}
\caption[] {
Variation of the energy dissipation at the shock as a function of
shock location. (a) Accretion rate of the flow is chosen as
${\dot {M}}=0.01 \times 10^{-5}$. Curves are plotted for a set of
angular momentum starting form $\lambda = 3.46$ (left most curve) to
$\lambda=3.82$ (right most curve) in the interval of $d\lambda = 0.04$. 
(b) Accretion rate of the flow is chosen as ${\dot {M}}=0.01 \times 10^{-6}$. 
The leftmost curve is drawn for $\lambda = 3.14$ and other curves are
for increasing angular momentum with an interval of $d\lambda = 0.04$. 
}}
\end{figure}

Fig. 1 shows a phase space trajectory of accretion flow around a
Schwarzschild black hole where the logarithmic radial distance is plotted
along the horizontal axis and the Mach number ($M =\vartheta/a$) is
plotted along the vertical axis. Arrows indicate the 
direction of the accreting flow. Here, the flow parameters are 
$\lambda=3.56$, ${\dot {\cal M}=0.08 \times 10^{-5}}$ and $\gamma = 4/3$.
Sub-sonic accreting flow passes 
through the outer sonic point (O) and becomes super-sonic. Depending 
on the energy dissipation at the shock $(\Delta {\cal E})$, shock 
conditions are satisfied at some particular location and the flow makes 
a discontinuous jump (shock-transition) from super-sonic branch to
sub-sonic branch there. After the shock transition, the flow momentarily
slows down and subsequently picks up it velocity to cross the inner sonic
point (I) before entering into the black hole. Since the flow losses a
part of its energy at the shock in the form of thermal energy, the pressure
in the post-shock region is diminished. As a result, the shock moves
forward to maintain a pressure balance across it. The possible shock
transitions at $x_s$ are shown as the vertical lines for various energy
dissipations at the shock.  For a given set of input parameters
$(\lambda =  3.56, {\dot {\cal M}} = 0.08 \times 10^{-5})$, the shock
forms closer to the black hole horizon as the energy dissipation at the
shock increases. For instance, $x_s = 27.75,~~\Delta {\cal E} 
= 0.699 \times 10^{-3}$ (big-dashed-dotted), $ x_s = 17.02,~~ \Delta 
{\cal E}=0.37 \times 10^{-2}$ (small-dashed-dotted)
and $x_s = 10.0,~~ \Delta {\cal E} = 0.985 \times 10^{-2}$
(dotted) respectively.  

\begin{figure}%[htb]
\vbox{
\vskip -0.3cm
\centerline{
\psfig{figure=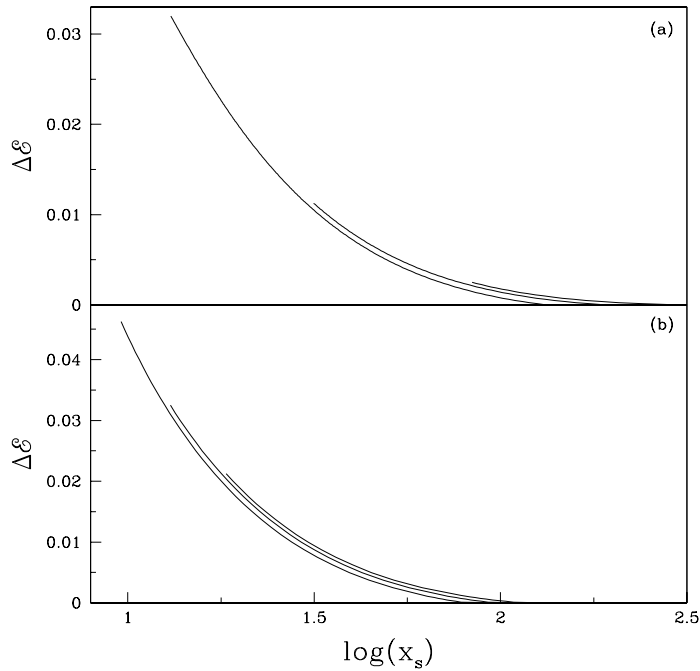,height=10truecm,width=10truecm}}
\vspace{-0.0cm}
\caption[] {
Plot of the energy dissipation at the shock with the shock location for
a given angular momentum. (a) $\lambda = 3.70$ and (b) $\lambda = 3.22$. 
See text for more details.
}}
\end{figure}

In Fig. 2(a-b), we present the variation of shock location 
in the logarithmic scale as a function of the energy dissipation at
the shock. In the upper panel, we show the results obtained by using
PW potential while in the lower panel, we use the PK potential.
In Fig. 2a, the accretion rate is chosen as ${\dot 
{\cal M}} = 0.01 \times 10^{-5}$ and adiabatic index is $\gamma=4/3$.
Different curves are drawn for a set of angular momentum starting from
$\lambda = 3.46$ (left most curve) to $\lambda = 3.82$ (right most curve)
in the interval of $d\lambda=0.04$. In Fig. 2b, we consider 
${\dot {\cal M}} = 0.01 \times 10^{-6}$ and black hole rotation 
parameter $a_k= 0.5$. In this plot, the angular momentum is varied from 
$\lambda = 3.14$ (left most curve) to $\lambda = 3.30$ (right most curve) 
in the interval of $d\lambda=0.04$. Both the Figures indicate that the
shock forms closer to the black hole horizon as the energy dissipation
increases monotonically for a given input parameters
$({\dot {\cal M}}, \lambda)$. The 
conclusion is valid for both the rotating and non-rotating black holes.
Note that there is a cut-off in each individual curve. At very high energy 
dissipation RHCs are not satisfied and no steady shocks can form.
However, non-steady shocks may still form as accreting matter continues to 
possess multiple sonic points where entropy at the inner sonic
point is higher than the outer sonic point \citep{dcc01}.
In this case, shock location becomes imaginary and shock starts oscillating
with a time period which is comparable to the inverse of observed quasi
periodic oscillation (QPO) frequencies of black hole candidates. 
In order to investigate this
physical process explicitly, a rigorous time dependent analysis
is required which is beyond the scope of our present work.
As the angular momentum is increased, maximum energy dissipation at
the shock first increases, becomes maximum at some $\lambda$ and then
decreases. For a non-rotating black hole, the maximum energy dissipation at
the shock ($\Delta {\cal E}_{max}$) is around 3.5\% for accretion 
flows with angular momentum $\lambda = 3.70$ and for rotating black 
hole $\Delta {\cal E}_{max}$ is around 4.7\% for $\lambda = 3.22$. 

A diagram similar to Fig. 2(a-b) is drawn in Fig. 3(a-b) for various
entropy accretion rates keeping the specific angular momentum fixed.
In the upper panel, we present the
results for flows with angular momentum $\lambda = 3.70$ accreting 
onto a non-rotating black hole. Here, ${\cal M}$ is varied from 
$\dot {\cal M}= 0.01 \times 10^{-5}$ (leftmost) to $0.21 \times 
10^{-5} $ (rightmost) in the interval of $d{\dot {\cal M}} = 0.01 
\times 10^{-5}$. In the lower panel, we present a similar plot 
for flows of angular momentum $\lambda = 3.22$ accreting around
a rotating black hole with the rotation parameter $a_k= 0.5$. The 
accretion rate is chosen to be $\dot {\cal M}= 0.01 \times 10^{-6}$ 
(left most curve) to $0.16 \times 10^{-6} $ (right most curve) in the interval 
of $d{\dot {\cal M}} = 0.05 \times 10^{-6}$. In both the Figures,
for a given accretion rate, the shock front moves forward with the 
enhancement of energy dissipation at the shock. Interestingly,
there is a significant reduction in maximum energy dissipation 
($\Delta {\cal E}_{max}$) at the shock as ${\cal M}$ is 
increased. Thus the parameter space for standing shock is reduced with
the increase of the energy dissipation at the shock.

\begin{figure}%[htb]
\vbox{
\vskip -0.3cm
\centerline{
\psfig{figure=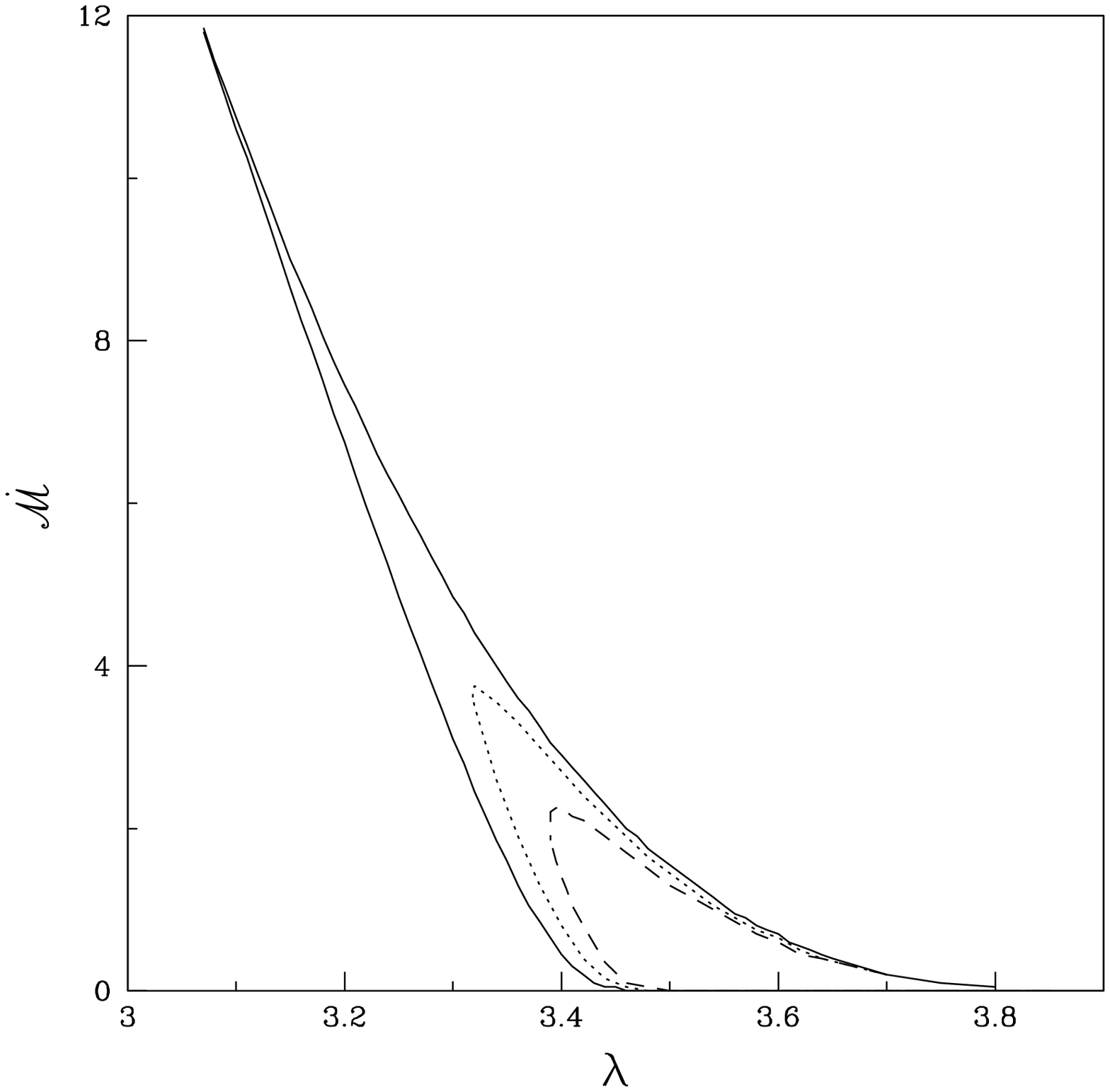,height=8truecm,width=8truecm}
\hskip -1.0cm
\psfig{figure=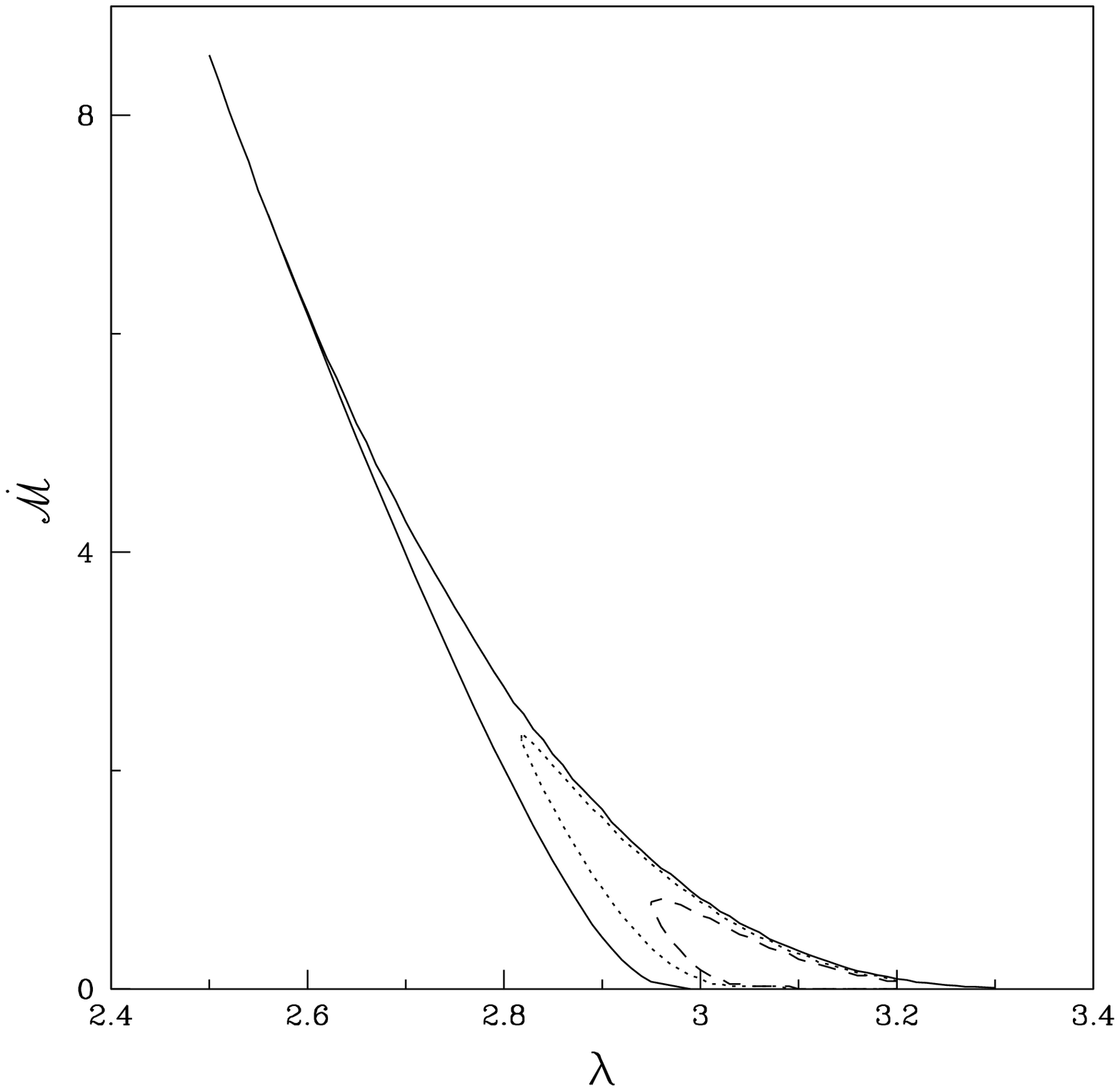,height=8truecm,width=8truecm}}
\vspace{-0.0cm}
\caption[] {
Modification of the region of the parameter space which forms a standing shock 
as a function of energy dissipation $(\Delta {\cal E})$ at the shock.
Left panel: For non-rotating black hole. Right panel: For rotating black
hole ($a_k=0.5$). Parameter space shrinks with the increase of
$\Delta {\cal E}$.
}}
\end{figure}

In Fig. 4, we classify the parameter space spanned by the accretion
rate $(\dot {\cal M})$ and the angular momentum $(\lambda)$ on the 
region which allows standing shocks for various values of energy
dissipation at the shock. In the left and the right panels, the parameter 
spaces are obtained for non-rotating and rotating black holes. Here,
rotating black hole has $a_k =0.5$. 
In the left Figure, the parameter space under the solid curve
is obtained for a dissipation-free accretion flow.
The region under the dotted and the dashed curves are obtained for 
$\Delta {\cal E} =1.5 \times 10^{-3}$ and $3.0 \times 10^{-3}$ 
respectively. Similarly, in the right panel, the parameter spaces 
separated by solid, dotted and dashed curves are obtained for
$\Delta {\cal E} = 0.0, 1.0 \times 10^{-3}$ and $4.0 \times 10^{-3}$. 
These figures show that the effective region of the parameter space 
shrinks with the increase of the energy dissipation at the shock
in the lower angular momentum side. Above a critical limit of 
$\Delta {\cal E}$, this region disappears completely.

\begin{figure}%[htb]
\vbox{
\vskip -0.3cm
\centerline{
\psfig{figure=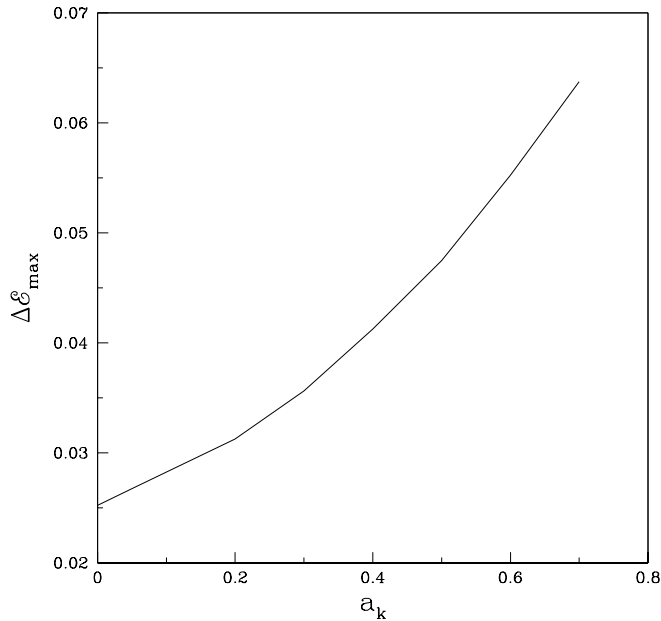,height=10truecm,width=10truecm}}
\vspace{-0.0cm}
\caption[] {Variation of $\Delta {\cal E}_{max}$ with black hole
rotation parameter $a_k$. See text for more details. 
}}
\end{figure}

An important part of understanding an accretion flow around a rotating
black hole is to compute the maximum energy dissipation 
($\Delta {\cal E}_{max}$) at the shock as a function of the black hole 
rotation parameter $a_k$. In Fig. 5, we present the variation of 
$\Delta {\cal E}_{max}$ with $a_k$. For instance, for an weakly rotating 
black hole, $\Delta {\cal E}_{max}$ is around 2.5\%. For increasing
$a_k$, $\Delta {\cal E}_{max}$ increases and reaches upto 6.5\% for a
rapidly rotating black hole. However, the maximum
available energy is around $5.7$\% and $13$\% of the rest mass
energy for $a_k=0$ and $a_k=0.8$ respectively. Thus we find that maximum
released energy at the shock is about $40-50$\% of the available energy.

\section{Conclusions}

It is well known that the accretion flows around a stellar mass black hole
emit hard X-rays from a Compton cloud located close to the inner edge of
the disk \citep{ct95}. This Compton cloud is the post-shock region which
also produces jets and outflows from the disk. The Comptonization process
removes mainly the thermal energy of the flow. In this paper, we show that
at standing shock the maximum release of energy could be as high as
$40-50$\% of the maximum available energy. Since the radiative loss
removes a significant thermal pressure from the inflow, the location
of the shock is affected. In particular, the shock seems to form closer
to the black hole when the dissipation is increased due to the drop of the
post-shock pressure which no longer can support the shock in its
original place. The implication of this could be profound. First of
all, the observed QPOs are considered to be due to the oscillation
of this shock.  During outbursts, the QPOs of several black hole
candidates, such as, XTE J1550-564 and GRO J1655-40 evolve rapidly.
The frequency changes from several mHz to a few tens of Hz in the rising 
phase of the outbursts. Since the QPO frequency is inversely proportional
to the infall time of the post-shock flow, such an evolution indicates
that the shocks actually propagated towards the black hole at the rising
phase of the outbursts \citep{skc08,skc09}. Our mechanism discussed
here gives a plausible explanation of such a phenomenon.

\section*{Acknowledgments}

SKCs work is partly supported by a RESPOND project and SD is partly supported
via a postdoctoral fellowship from the Korea Astronomy and Space Science
Institute (KASI).


\begin{thebibliography}{}

\bibitem[\protect\citeauthoryear{Chakrabarti}{1989}]{c89}
Chakrabarti S. K., 1989, ApJ 347, 365.

\bibitem [\protect\citeauthoryear{Chakrabarti}{1990}]{c90}
Chakrabarti S. K., 1990, Theory of Transonic Astrophysical Flows.
World Scientific Publishing, Singapore.

\bibitem[\protect\citeauthoryear{Chakrabarti \& Titarchuk}{1995}]{ct95}
Chakrabarti S. K., Titarchuk L. G., 1995,  ApJ, 455, 623.

\bibitem[\protect\citeauthoryear{Chakrabarti}{1999}]{c99}
Chakrabarti, S. K., 1999, A\&A 351, 185.

\bibitem[\protect\citeauthoryear{Chakrabarti \& Das}{2004}]{cd04}
Chakrabarti S. K., Das S., 2004, MNRAS, 349, 649. 

\bibitem[\protect\citeauthoryear{Chakrabarti \& Mondal}{2006}]{skcsm06}
Chakrabarti S. K., Mondal S., 2006, MNRAS, 369, 976.

\bibitem[\protect\citeauthoryear{Chakrabarti et al.}{2008}]{skc08}
Chakrabarti, S. K. Debnath, D., Nandi, A., Pal, P. S., 2008, A\&A,
489, 41.

\bibitem[\protect\citeauthoryear{Chakrabarti et al.}{2009}]{skc09}
Chakrabarti, S. K., Datta, B. G. Pal, P. S., 2009, MNRAS (in press).

\bibitem[\protect\citeauthoryear{Das et al.}{2001}]{dcc01}
Das S., Chattopadhyay I., Chakrabarti S. K., 2001, ApJ, 557, 983.

\bibitem[\protect\citeauthoryear{Das et al.}{2001}]{dcnc01}
Das S., Chattopadhyay I., Nandi, A., Chakrabarti S. K., 2001, A\&A, 379, 683.

\bibitem[\protect\citeauthoryear{Das \& Chakrabarti}{2004}]{dc04}
Das S., Chakrabarti S. K., 2004, IJMPD, 13, 1955.

\bibitem[\protect\citeauthoryear{Das}{2007}]{d07}
Das S., 2007, MNRAS, 376, 1659.

\bibitem[\protect\citeauthoryear{Das \& Chattopadhyay}{2008}]{dc08}
Das S., Chattopadhyay I., 2008, New Astronomy, 13, 549 .

\bibitem[\protect\citeauthoryear{Landau \& Lifshitz}{1959}]{ll59}
Landau L. D., Lifshitz E. D., 1959, Fluid Mechanics, New York: Pergamon.

\bibitem[\protect\citeauthoryear{Matsumoto et al.}{1984}]{mk84}
Matsumoto R., Kato S., Fukue J., Okazaki A. T., 1984, PASJ, 36, 7.

\bibitem[\protect\citeauthoryear{Paczynski \& Wiita}{1980}]{pw80}
Paczynski, B. and Wiita, P.J., 1980, Astron. Astrophys., 88, 23.


\end{thebibliography}
\end{document}